\documentclass{cjaa}                    

\input{psfig.sty}                       
 
\newcommand{\bq}{\begin{equation}}
\newcommand{\eq}{\end{equation}}
\begin{document}
\title{Color-redshift relations
and photometric redshift  estimations of quasars in large sky surveys}

\author{Xue-Bing Wu \inst{1},  Wei Zhang \inst{1} \& Xu Zhou \inst{2}}
\institute{1. Department of Astronomy, School of Physics, Peking University, Beijing
100871, China\\
2. National Astronomical Observatories, Chinese Academy of Sciences, Beijing 100012, China}

\offprints{wuxb@bac.pku.edu.cn}

\abstract
{
With a recently constructed composite quasar spectrum and
the $\chi^2$ minimization technique, we demonstrated a general method to estimate the 
photometric redshifts of
a large sample of quasars by deriving the theoretical color-redshift relations and 
comparing the  theoretical colors with the observed ones.
We estimated the photometric redshifts from the 5-band SDSS photometric data of
 18678 quasars in the first major data release of SDSS and compare them with the spectroscopic
redshifts.  The redshift difference is smaller 
than  0.1 for 47$\%$ of quasars and 0.2 for  68 $\%$ of 
them. Based on the calculation of the theoretical color-color diagrams of
stars, galaxies and quasars in both the SDSS and BATC systems, we expected that
with BATC intermediate band photometric system we would be able to select candidates of 
high redshift  quasars
more efficiently than in the SDSS, provided the BATC survey could detect 
objects with magnitude fainter than 21.
}

  \authorrunning{ X.-B. Wu, W. Zhang, \& X. Zhou}            
   \titlerunning{Color-redshift relations and photometric redshifts of quasars }   
 
   \maketitle

\section{Introduction}
\label{Sect1}

Quasars are intrinsically very luminous objects with absolute magnitude
brighter than $\rm{M_V}<-23$, but many of them looks very faint because they are usually 
very far away from us. The morphology of quasars looks like stars on the photographic and 
CCD plates
and we can not identify them directly from the images.
However, the spectrum of a typical quasar usually consists of strong broad
emission lines and a power law continuum (Francis et al. 1992; Vanden Berk 
et al. 2001), which is
quite different from stars.
Therefore it is relatively easier to separate quasars from stars with the spectroscopic 
observations.
In recent years, more and more quasars have been discovered along
with the ongoing projects of large sky surveys such as SDSS
( Sloan Digital Sky Survey), 2dF (2 degree Fields)  and BATC (Beijing-Arizona-Taipei-Connecticut).
Many of these quasars have large redshift and therefore are very much helpful in studying the structure and evolution of the early universe.

Astronomical photometry can be used as low-resolution
spectroscopy (Bessell 1990). Photometric observation has its advantage of surveying 
a large sample of objects in large sky
areas. It can be used to identify much fainter 
objects than the 
spectroscopic observations at the same exposure time. The practice
of measuring redshift of galaxies and quasars using multi-color
photometry has become both very popular and powerful in recent years (Brunner et al. 1997; Connolly et al. 1999; Xia et al. 2002). Among many current sky survey programs with multi-color photometry, 
SDSS is a joint project which aims to discover more than $10^6$ galaxies and
$10^5$ quasars (York et al. 2000). The photometric survey in SDSS is being done with 5 
passbands( u', g', r', i', 
z') covering from from $3000 \AA$ to $10500 \AA$ (Fukugita et al. 1996).
Such a 5-band photometry can be treated as an $R\sim 4$
objective-prism survey (Richards et al. 2001). With these photometric data, quasar 
candidates can be efficiently
selected with some criteria (Fan 1999; Newberg et al. 1999; Richards et al. 2002). 
Subsequent spectroscopic 
observations of these 
candidates
have revealed 18678 quasars in the first major data release of SDSS (Abazajian et al. 2003).
Discovery of several quasars with redshift larger than 6 in the SDSS project has given
us important information about the reionization epoch of our universe (Becker et al. 2001;
Fan et al. 2002).

In China, we have also developed a program of multi-color photometric survey, namely 
the BATC project (Chen 1998; Zhou et al. 2001; Zhou et al. 2003). This is being done with a 
60/90cm Schmidt telescope at the National Astronomical Observatories of Chinese Academy of Sciences. A Ford Aerospace CCD with $1024\times 1024$ pixels and 15
intermediate-band filters covering from 3200\AA~ to 10000\AA~ were used in the Survey.
The field of view of the telescope is about 1 deg$^2$ and the limit magnitude for 
photometry  reaches 21 mag approximately in V band. 
Because the number of intermediate-band filters of
this system is much more than others, theoretically it
can obtain much more accurate photometric data such as the spectral energy distributions 
(SED) of many different types of
objects.
The survey started to work in 1995  and has so far surveyed
about 100 deg$^2$ sky region at high galactic latitudes. It has got many 
important results on
star clusters, nearby galaxies and galaxy clusters (Kong et al. 2000;  Yuan  et al. 2001;
Ma et al. 2001; Xia et al. 2002).

With the 15-band photometric data from the BATC survey, it could be more efficient to 
select quasar candidates than in other broad band surveys. In this paper, we first derive the theoretical color-redshift relations for quasars based on the composite spectrum obtained in the SDSS, and then use them to estimate the photometric redshifts of a large sample of quasars and compare them with the spectroscopic redshifts. Furthermore, we give the
theoretical color-color relations of stars, galaxies and quasars in both the 
SDSS and BATC surveys, and demonstrate that the BATC system
should be more efficient in selecting 
candidates of high redshift quasars than the SDSS, provided it can reach the 
magnitude fainter than 21.

\section{Color-redshift relations of quasars in the SDSS system}

\begin{figure}
\centering
\psfig{figure=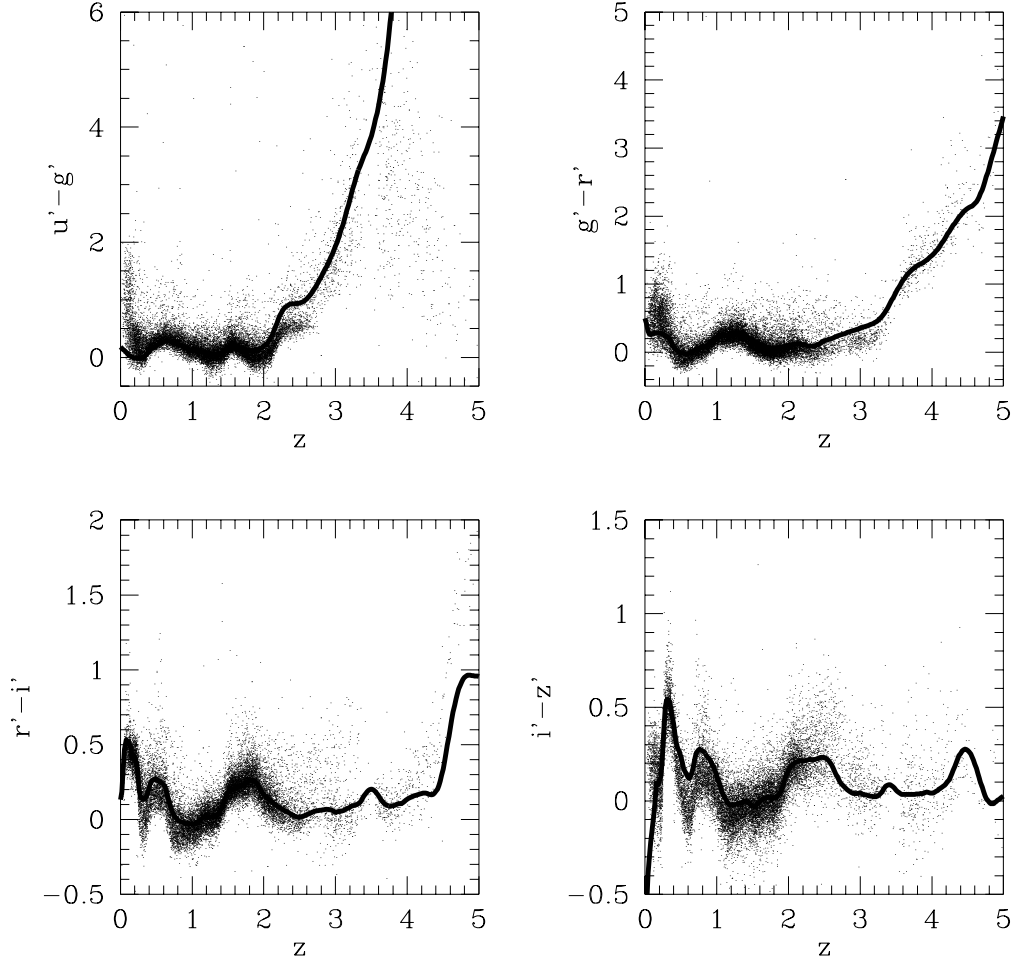,width=14cm}
\vskip -0.3cm
\caption{Theoretical color-redshift relation for quasars in the SDSS system (shown as solid lines) 
and its comparisons with the observation data of 18678 quasars (shown as dots) in the first major data release of SDSS.}
\end{figure}

Using the spectra for more than 2200 quasars in
the early data release, the SDSS team has derived a composite
spectrum for quasars (Vanden Berk et al. 2001). The wavelength range of this 
spectrum covers from 800 $\AA$ to 8550$\AA$ and is much broader than any previous
composite spectra of quasars (Francis et al. 1991; Zheng et al. 1997).
Assuming that this composite spectrum is the best representative for all quasars,
we can calculate the color-redshift relations for quasars in different photometric systems. 
This means, however,  that we have to omit other types of quasars such as quasars with 
broad absorption lines and redder colors in our calculations.

Under these assumptions, we can estimate the magnitude of quasars in every passband at different
redshift theoretically. If we define  $S(\nu)$ as the transmission efficiency of the
filter at frequency $\nu$, $f(\nu)$ as the flux of the quasar at the same frequency.
The flux of the quasar in this passband is expressed as:
\bq
f=\int_{\nu_1}^{\nu_2}f{(\nu)}S{(\nu)}{\rm d}\log{\nu}
\eq
Because SDSS system is an  AB photometry system (Fukugita et al. 1996), then the magnitude in this passband is:
\bq
m=-2.5\log{\frac{\int_{\nu_1}^{\nu_2}f{(\nu)}S{(\nu)}{\rm d}\log{\nu}}{\int_{\nu_1}^{\nu_2}S{(\nu)}{\rm d}\log{\nu}}}-48.60
\eq
Then the color of this quasar can be expressed as (Richards et al. 2001):
\bq
m_1-m_2=-2.5\left(\log{\frac{\int_{\nu_1}^{\nu_2}f{(\nu)}S_1{(\nu)}{\rm d}\log{\nu}}{\int_{\nu_1}^{\nu_2}S_1{(\nu)}{\rm d}\log{\nu}}}-
\log{\frac{\int_{\nu_3}^{\nu_4}f{(\nu)}S_2{(\nu)}{\rm d}\log{\nu}}{\int_{\nu_3}^{\nu_4}S_2{(\nu)}{\rm d}\log{\nu}}}\right)
\eq

\begin{figure}
\centering
\psfig{figure=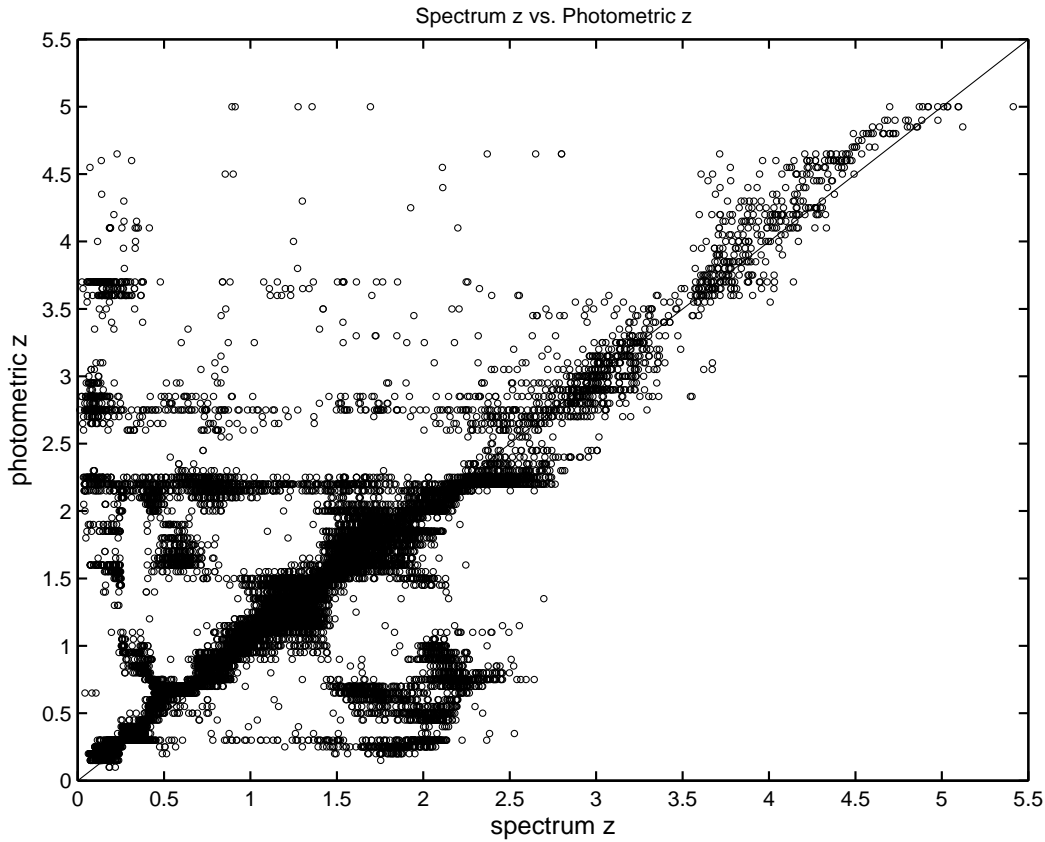,width=14cm}
\caption{Comparison of photometric redshifts with spectroscopic redshifts of 18678 SDSS quasars.
 The diagonal line shows the equal redshifts}
\end{figure}

With the transmission curves of SDSS filters and the composite quasar spectrum,
 we calculated the magnitudes in 
5 SDSS photometric bands at different redshift $z$, with a redshift bin of 0.05. From these magnitudes we can get 4 theoretical color-redshift relations of 
quasars in SDSS survey (shown as solid lines in Figure 1). 
We noted that  we can not get
reasonable value of the color $u'-g'$ at larger redshift. This is because Ly$\alpha$ emission line 
moves out of the response of filter $u'$ when 
redshift is larger than 3.6.
In addition, we found that our color-redshift relations are very similar as those obtained 
in Fan (1999) by assuming a power law continuum and the typical emission line ratios of quasars. However, the method we adopted to derive these relations is more straightforward than the previous one.

We also compared the theoretical color-redshift relations with the observational data of 18678 quasars in the first major data release of SDSS (Abazajian et al. 2003). The points in Figure 1 represent the
observation data of quasars. It is clear that our
theoretical results are well consistent with the observed ones.

\section{Estimation of the photometric redshifts of quasars}

The photometric redshifts of quasars can be estimated by comparing the observed colors with the theoretical 
color-redshift relations. A standard
$\chi^2$ minimization method is used here to estimate the most possible photometric redshifts of quasars. If we write the theoretical color as $m_{i,theory}-m_{j,theory}$, the observed
color as $m_{i,observed}-m_{j,observed}$ and the uncertainties of observed magnitudes in $i$ and $j$ band as $\sigma_{m_{i, observed}}$, $\sigma_{m_{j,observed}}$, the $\chi^2$ is then
defined as:
\bq
 \chi^2=\sum_{u',g',r',i',z'}{\frac{[(m_{i,theory}-m_{j,theory})-(m_{i,observed}-m_{j,observed})]^2}{\sigma_{m_{i, observed}}^2+\sigma_{m_{j,observed}}^2}}
\eq
Because SDSS photometric data have uncertainties  in 5 filter 
bands so we perform the $\chi^2$ minimization for all  4
colors.
For  a quasar we can get a sequence of $\chi^2$ for every different redshift between 0 and 
5.  If we further consider that all the
quasars should have absolute magnitude brighter than -23, this requirement will help us to
exclude
some results with extremely lower redshifts.

\begin{figure}
\centering
\psfig{figure=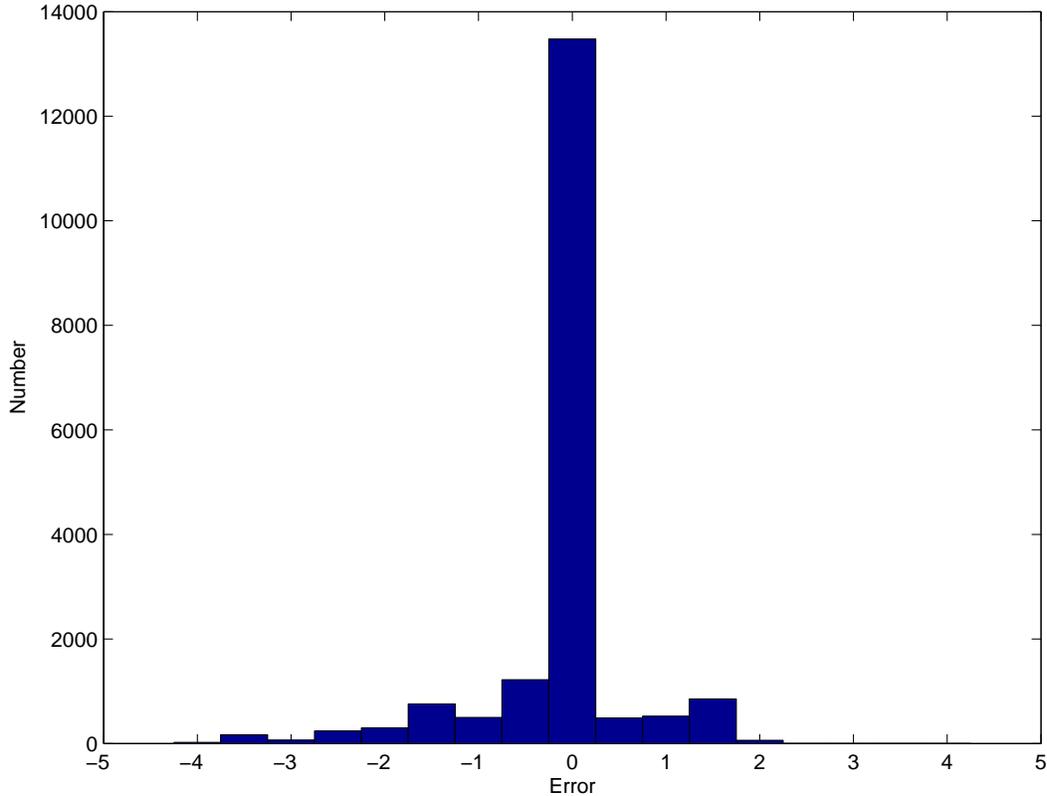,width=14cm}
\caption{Difference distributions between photometric and spectroscopic redshifts.}
\end{figure}

Following these steps, we can get the smallest $\chi^2$ and the corresponding 
photometric redshift for a quasar with 5-band photometric data. With this method we have 
calculated 
the photometric
redshifts for all 18678 quasars in the first major data release of SDSS. 
We noted that  Ly$\alpha$ emission line moves out of the response of filter $u'$ when 
redshift is larger than 3.6, so 
we would get
unreasonable value of the color $u'-g'$ at larger redshift. This means that we 
can only use 3 of 4 colors if the trying redshift is larger than 3.6.  However,  
we still used 4 colors to estimate the photometric redshift if the trying redshift
 is smaller than 3.6 in order to get more accurate $\chi^2$.

In the first major data release of SDSS, the  spectroscopic redshift of all the 18678 
quasars
have been given (Abazajian et al. 2003). With these redshifts, we can  check
the efficiency of our method. The comparison of the photometric redshifts and the 
spectroscopic redshifts is shown
in Figure 2. We found that the difference between photometric redshift and spectral redshift 
is smaller than 0.1 for 46.62\% of quasars and is smaller than 0.2 for 67.87\% of quasars. 
The histogram of these differences is shown in Figure 3. From it we can clearly see
that the photometric redshifts estimate by us are quite accurate within $|\Delta z| \le 0.2$
for most of quasars.

\begin{figure}
\centering
\psfig{figure=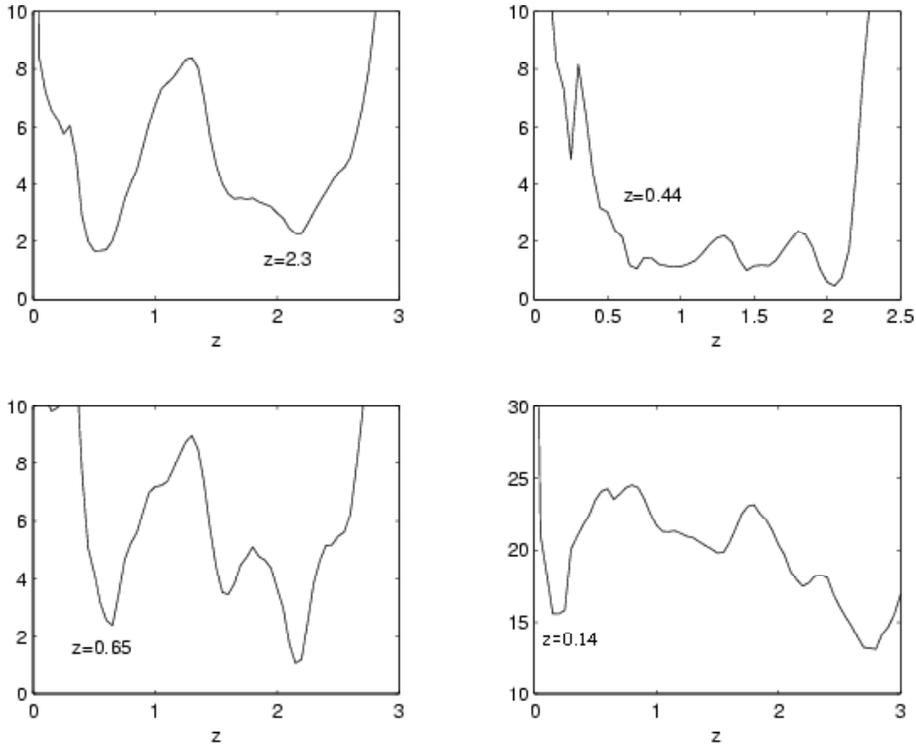,width=14cm}
\caption{Four examples of wrong photometric redshifts estimated using the $\chi^2$ minimization
method. The spectroscopic redshifts for these quasars are shown in the figures.}
\end{figure}

However, using
the $\chi^2$ minimization method to estimate the photometric redshift can also lead wrong results for some quasars. If the real spectra of quasars are substantially different from the composite spectrum or the uncertainties of photometric magnitudes are larger, our calculated $\chi^2$ might reach the minimum at a wrong
redshift. Some examples of these cases are shown in Figure 4.
From them we can clearly see that the $\chi^2$ gets minimum value at wrong photometric redshift. Such a problem could be alleviated if more composite spectra of different types of quasars were adopted to calculate the theoretical color-redshift relations.

We noticed that the SDSS team has also made the estimations of photometric redshift of 2625
 quasars in the early data release of SDSS (Richards et al. 2001). However, they adopted a different method
from that adopted by us to derive the color-redshift relation. As described in Richards et al. (2001),
 their color-redshift relation was the median of 2200 quasars in the early data release of SDSS
 They
calculated the photometric redshift for these 2625 quasars by using the 
$\chi^2$ minimization method 
with these median color-redshift relations
and the observational colors. The difference between their photometric redshift and 
spectral redshift is smaller than 0.1 for 55\% of quasars
and  is smaller than 0.2 for 70\% of them.
Although the efficiency of our method for photometric redshift estimation
is slightly lower than that adopted by the SDSS team, the major advantage of
our method is that it can be easily adapted to any other photometric systems.
With our method, the theoretical color-redshift relation of quasars in other photometric systems
can be easily obtained as long as the transmission curves of the filters are given. In contrary, 
the median color-redshift relation for quasars got by the SDSS team can only be used for 
the SDSS photometric data.
In addition, we also noted that the efficiency of our photometric redshift
estimation is comparable to that in some other groups such as 
Hatziminaoglou et al. (2000) and  Wolf et al. (2001), who estimated the 
photometric redshifts of smaller samples of quasars with different methods.

\section{Color-color diagrams of quasars, stars and galaxies}

\begin{figure}
\centering
\psfig{figure=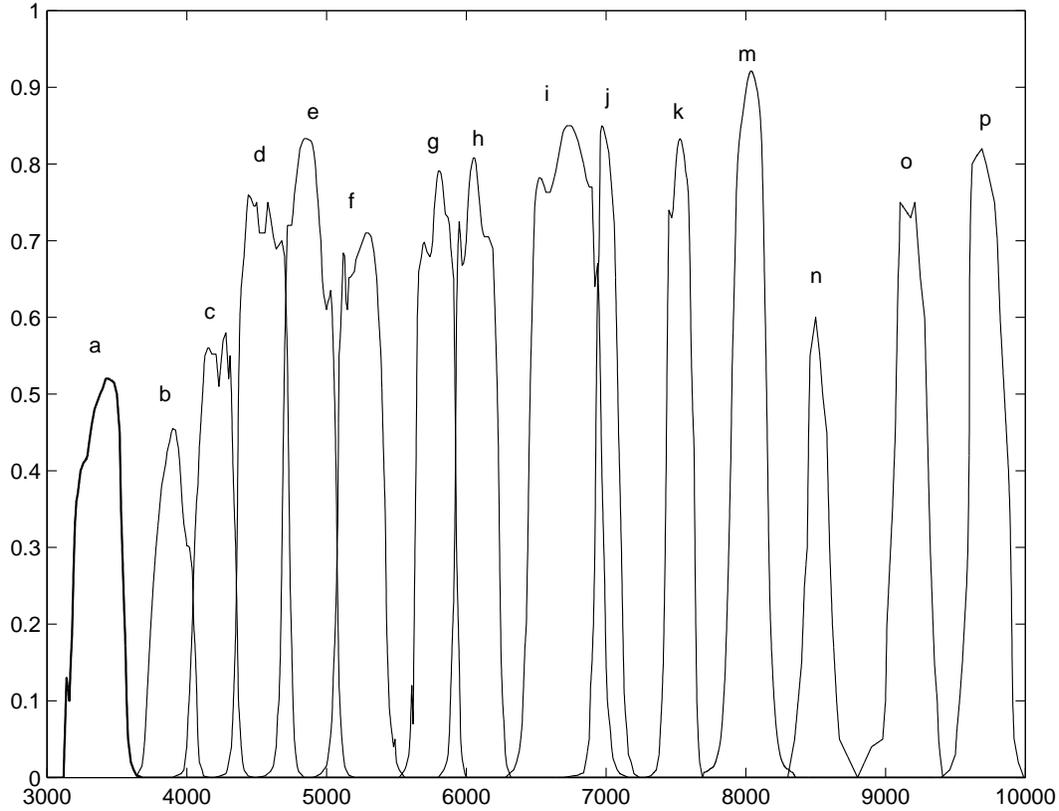,width=14cm}
\vskip 0.2cm
\caption{The transmission curves of 15 intermediate band filters in BATC survey. The names of each filters are also shown on the top of the curves.}
\end{figure}

\begin{figure}
\centering
\psfig{figure=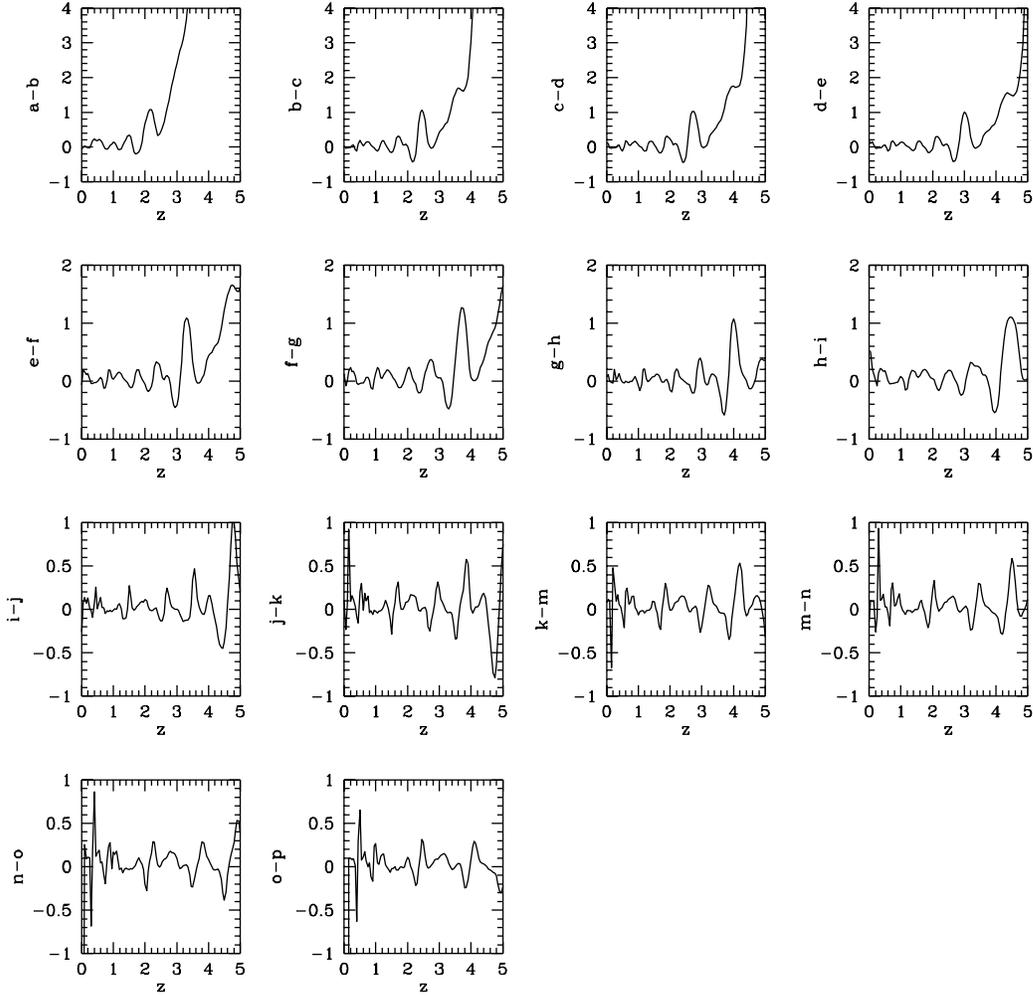,width=15cm}
\vskip -0.4cm
\caption{The theoretical color-redshift relations of quasars in the BATC system.}
\end{figure}

We used the method described above to calculate the theoretical colors of quasars, stars and galaxies in the BATC system in order to find some  criteria for
selecting quasar candidates from the survey data. 
The transmission curves of BATC 15 intermediate filters (named from $a$ to $p$ except $l$) 
are shown in Figure 5 
(see also Xia et al 2002). 
We did the calculation of 14 color-redshift relations
of quasars in the BATC system using the composite quasar spectrum obtained by the SDSS. 
The results are shown  in Figure 6.
For stars, we used  the stellar spectral library given by Pickles et al. (1998), which 
consist 130 template spectra of stars with different spectral types from O to M. 
For galaxies, we used the template spectra of 12 types of galaxies (Kinney et al. 1996). 
We calculated
the colors of each type of galaxy at different redshift from 0 to 5 with a bin of 0.05.

\begin{figure}
\centering
\psfig{figure=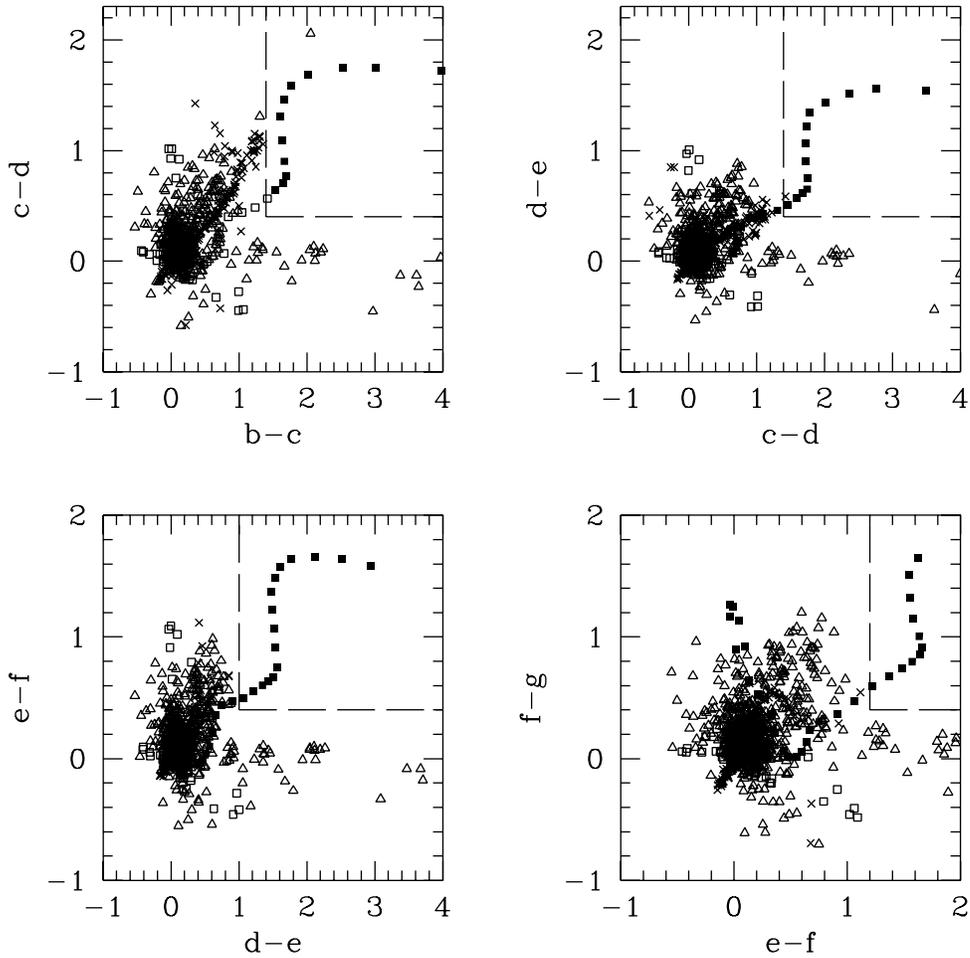,width=15cm}
\vskip -0.9cm
\caption{Four theoretical color-color diagrams of stars, galaxies and quasars in the BATC system.
The crosses denote stars and the triangles denote galaxies. The filled and open squares 
denote quasars with redshift larger and smaller than 3.5 respectively. The right upper regions separated by dashed lines in
each diagrams are dominated by high redshift quasars.}
\end{figure}

\begin{figure}
\centering
\psfig{figure=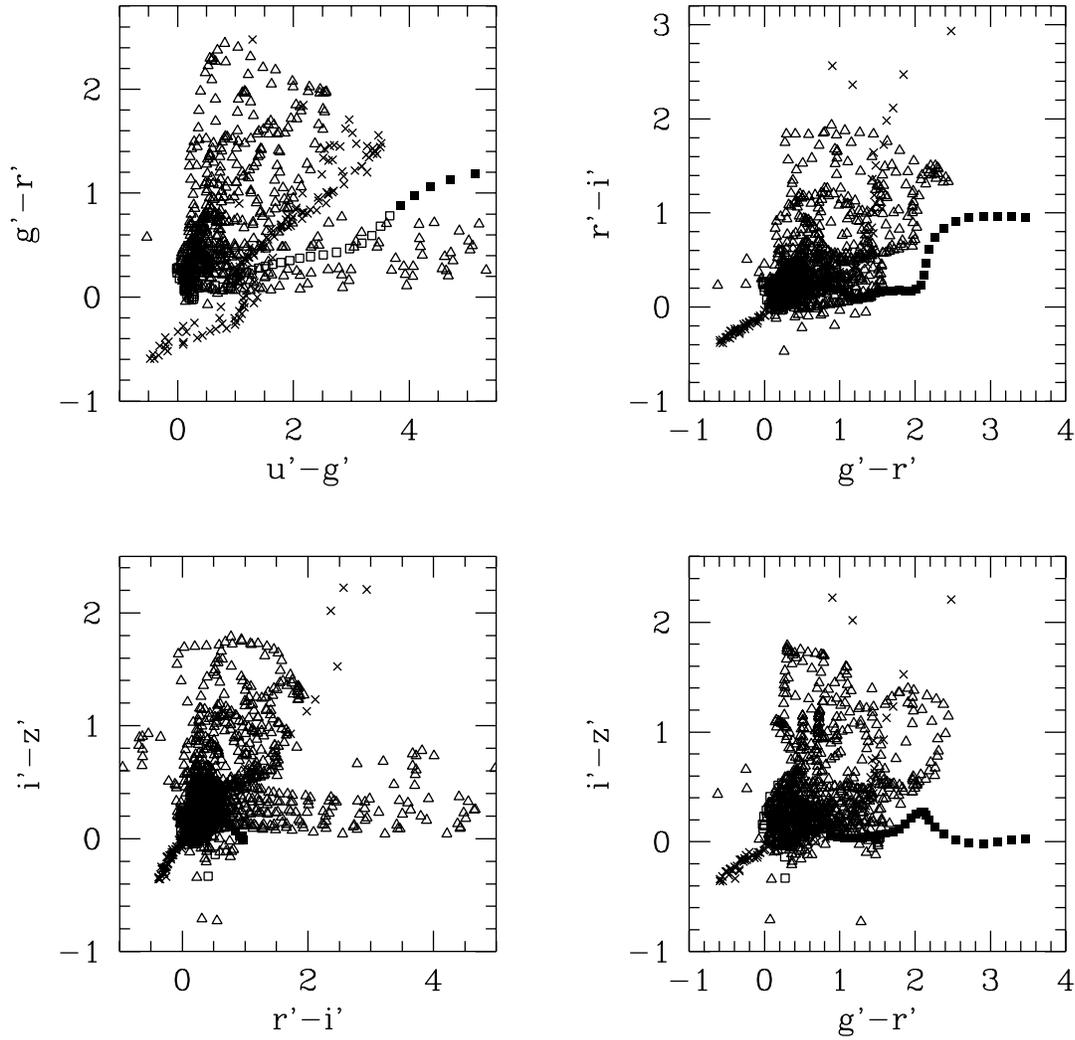,width=15cm}
\vskip -0.2cm
\caption{Four theoretical color-color diagrams of stars, galaxies and quasars in the SDSS system.
The crosses denote stars and the triangles denote galaxies. The filled and open squares 
denote quasars with redshift larger and smaller than 3.5 respectively.}
\end{figure}

We have plotted stars, galaxies and quasars on 13 color-color diagrams and found that it 
is not easy to use a simple criterion to separate quasars from stars in many of these diagrams. 
But it is still possible to separate quasars (especially high redshift ones) from stars 
and galaxies on 4 of the color-color
diagrams (shown in Figure 7). 
We noted that most stars
lie in a narrow belt on the color-color diagrams, while
galaxies with different redshifts distribute much diversely.
However, it is clear that the right upper regions in
these diagrams are dominated by high redshift quasars. Therefore the dashed lines
in these 4 diagrams can be used as the simple criteria to select candidates of quasars 
with redshift larger than 3.5 from the photometric data in the BATC survey. These criteria are:
$$
b-c >1.4~ and~ c-d >0.4;$$ $$ or~ c-d>1.4~ and~ d-e>0.4;$$$$ or~ d-e>1~ and~ e-f>0.4; 
$$ $$ or ~e-f>1.2~ and~ f-g>0.4 $$ 

In addition, we note that quasars 
are mostly point sources while many of
galaxies are extended objects. Therefore we can further exclude some galaxies using morphology 
information. After these checks,
candidates of high redshift quasars can be then selected from these color-color diagrams for further 
spectroscopic identifications. 
We also noted that quasar candidates selected in this way may be still contaminated 
with 
some brown dwarfs and some other kind
of AGNs (Fan et al. 2002). Further detailed studies on the differences of brown dwarfs and 
high redshift quasars in the color-color diagrams may help us to solve this problem.

We note that with our selection criteria we will still miss  a lot of quasars 
, especially those with
lower redshift because they  locate at the same places as stars in the color-color
diagrams. But higher efficiency of selecting high redshift quasars from these color-color 
diagrams is much more 
attractive and this could be an
advantage of BATC photometric system because it has 15 intermediate filters and is able to identify the 
main emission lines more accurately than in other broad band photometric surveys.

Following the same method as above,
we also calculated the color-color diagrams of stars, galaxies and quasars in SDSS 
system.  The results are shown in
Figure 8, which is also similiar as those got by Fan (1999). Comparing with 
those in the BATC system, we found that the separation of quasars from stars and
galaxies on the diagrams in the SDSS system is not so well as in the BATC system. 
Although the candidates of high redshift quasars can be possibly selected only from 
the $(r'-i')$ vs $(g'-r')$ and the $(i'-z')$ vs $(g'-r')$ diagrams, the contaminations of stars and
galaxies
to the selection of quasar candidates are much more serious than in the BATC system.
This 
is understandable because of the broader filter bands of SDSS photometric system. Therefore, with the 
BATC photometric data we could be able to find candidates of high redshift 
quasars more efficiently than in the SDSS survey. 

\section{Discussions}

In this paper we demonstrated a method to derive the color-redshift relations of quasars from the 
composite quasar spectrum and
applied this method to SDSS and BATC systems. Our estimated photometric redshifts of 18687 quasars 
are well consistent with the spectroscopic 
redshifts given in the first major data release of SDSS. This method can be easily adapted to other multi-color survey system because it only needs the transmission functions of the filters and the photometric data. By comparing 
the theoretical color-color diagrams of stars, galaxies and quasars in the SDSS and BATC systems, we
found the BATC system would be better than SDSS in selecting high redshift quasar candidates 
provided 
the BATC survey could go as deeper as SDSS to detect objects fainter than 21 magnitude. A detailed
comparison of photometric redshifts, estimated from both the SDSS and BATC photometric data, for 
quasars in several common sky fields in both surveys will be given in another paper.

We noted that all the calculations of photometric redshifts in this paper is based on the
composite spectrum of SDSS quasars. However, when we use this composite spectrum for all quasars, 
there must 
be big errors. In fact, the spectra of quasars are much diverse. For example, the continuum and emission line intensity  of radio 
loud quasars are different from those of radio quite ones (Francis et al. 1993; Zheng et al. 1997). The quasars with  broad 
absorption lines and redder quasars have very different spectra from normal ones(Sprayberry 
\& Foltz 1992; Richards et al. 2003). These 
differences should be considered in estimating the photometric redshift of quasars.
In our future work we will use more composite spectra of different types of quasars. In addition,
we note that the  calculation of minimum  $\chi^2$  could be further improved with the ASQ 
method 
developed recently by the SDSS team to give more accurate photometric redshift of quasars 
(Budavari et al. 2001).

Large  sky area surveys have become very popular in recent years. 
The huge amount of data obtained from these surveys such as SDSS and 2DF already provided us much 
information about the 
structure and evolution of the universe. Comparison with the spectroscopic survey,
multi-color photometric survey over large sky areas
can obtain the spectral energy distributions of much more fainter objects. 
With the technique of 
photometric
redshift determination, we can construct large samples of objects with estimated redshifts and make
many interesting studies on galaxies and quasars. 
The multi-color photometric data can be also used to estimate the spectral types of numerous stars.
Therefore, from the multi-color photometric 
observations, we can efficiently select candidates of galaxies, quasars and some specific stars 
and prepare the input catalog for the further spectroscopic
surveys. This is very important especially for LAMOST, a future spectrocopic survey project 
in China (Luo \& Zhao 2001).

\acknowledgements
This work is supported by the National Key Project on Fundamental Researches (TG 1999075403) and the 
National Science Fundation (No. 10173001) of China.

\end{document}